\documentclass[usenatbib,useAMS]{mn2e}
\usepackage{graphicx}
\usepackage{ctable}
\usepackage{url}
\usepackage{times}
\usepackage{xspace,subfigure}

\newcommand{\msun}{~\mathrm{M_{\odot}}}

\newcommand{\sunrise}{\textsc{sunrise}\xspace}

\newcommand{\gadgettwo}{\textsc{gadget-2}\xspace}

\newcommand{\magphys}{\textsc{magphys}\xspace}

\newcommand{\apj}{ApJ}
\newcommand{\apjl}{ApJL}
\newcommand{\apjs}{ApJS}
\newcommand{\aj}{AJ}
\newcommand{\apss}{Ap\&SS}
\newcommand{\mnras}{MNRAS}
\newcommand{\nat}{Nat}

\newcommand{\pasp}{PASP}
\newcommand{\araa}{ARA\&A}
\newcommand{\aap}{A\&A}

\newcommand\qjras{{QJRAS}}

\newcommand{\acknowledgments}{\begin{small}\section*{Acknowledgments}\end{small}}

\newcommand{\dM}{\ensuremath{\Delta M \slash M}}
\newcommand{\chisqsfh}{\ensuremath{\chi^2_{\rm SFH}}}

\newcommand\sref[1]{\hyperref[#1]{Section~\ref*{#1}}}
\newcommand\fref[1]{\hyperref[#1]{Fig.~\ref*{#1}}}

\title[Star formation histories using photometric SED modelling]{
Deriving star formation histories from photometry using energy balance spectral energy distribution modelling
}

\author[D.~J.~B. Smith \&\ C.~C. Hayward]{
\parbox[t]{\textwidth}{
Daniel J.~B.~Smith$^1$\thanks{E-mail: daniel.j.b.smith@gmail.com}
and 
Christopher C. Hayward$^{2,3}$
\thanks{Moore Prize Postdoctoral Scholar in Theoretical Astrophysics}
}
\vspace*{6pt} \\
$^1$Centre for Astrophysics, Science \& Technology Research Institute, University of Hertfordshire, Hatfield, Herts, AL10 9AB, UK\\
$^2$TAPIR 350-17, California Institute of Technology, 1200 E. California Boulevard, Pasadena, CA 91125, USA\\
$^3$Harvard--Smithsonian Center for Astrophysics, 60 Garden Street, Cambridge, MA 02138, USA
}

\usepackage[breaklinks,colorlinks,allcolors=blue]{hyperref}
\usepackage[all]{hypcap}
\usepackage{amssymb}

\begin{document}

\date{MNRAS {\it in press}}

\pagerange{\pageref{firstpage}--\pageref{lastpage}} \pubyear{2015}

\maketitle

\label{firstpage}

\begin{abstract}
Panchromatic spectral energy distribution (SED) fitting is a critical tool for determining the physical properties of distant galaxies, such as their stellar mass and star formation rate. One widely used method is the publicly available \magphys\ code.
We build on our previous analysis \citep{HS15} by presenting some modifications which enable \magphys\ to  automatically estimate
galaxy star formation histories (SFHs), including uncertainties, based on ultra-violet to far-infrared photometry.
We use state-of-the art synthetic photometry derived by performing three-dimensional dust radiative transfer on hydrodynamic simulations of isolated disc and merging galaxies to test
how well the modified \magphys\ is able to recover SFHs under idealised conditions, where the true SFH is known. We find that while the SFH of the
model with the best fit to the synthetic photometry is a poor representation of the true SFH (showing large variations with the line-of-sight to the galaxy and spurious bursts of star
formation), median-likelihood SFHs generated by marginalising over the default \magphys\ libraries produce robust estimates of the smoothly-varying isolated disk simulation SFHs. This preference for the median-likelihood SFH is quantitatively underlined by our estimates of \chisqsfh\ (analogous to the $\chi^2$ goodness-of-fit estimator) and \dM\ (the
integrated absolute mass discrepancy between the model and true SFH) that strongly prefer the median-likelihood SFHs over those that best fit the
UV-to-far-IR photometry. In contrast, we are unable to derive a good estimate of the SFH for the merger simulations (either best-fit or median-likelihood)
despite being able to obtain a reasonable fit to the simulated photometry, likely because the analytic SFHs with bursts superposed in the standard \magphys\ library are insufficiently general\slash realistic.
\end{abstract}


\section{Introduction} \label{S:intro}

Determining the star formation histories (SFHs) of galaxies is of
paramount importance for understanding galaxy formation and
evolution. For example, the SFHs of galaxies can reveal signatures of
interactions and yield insight into the physics of
feedback. Connecting galaxy populations at different epochs can help
elucidate the typical SFHs of galaxies, but it is difficult to
unambiguously determine the progenitors and descendants of a given
galaxy population \citep[though see e.g.][]{behroozi10,guo10,mundy15}.
For this reason, inferring the SFHs of
individual objects -- if it is possible to do so accurately -- would be preferred.
Moreover, reliable individual SFHs
for large numbers of galaxies would enable more detailed comparisons
with simulations than are currently possible. For example, simulated
and observed galaxy SFHs could be used to determine whether the
simulations reproduce the SFHs of real galaxies, not just the
statistical properties of galaxy populations
\citep[e.g.][]{Cohn15,Shamshiri15,Sparre15}.

When it is possible to resolve individual stars, one can determine a
galaxy's SFH from its colour-magnitude
diagram \citep[e.g.][]{Tosi89,Tosi91,Bertelli92,Tolstoy96,Hernandez99,Hernandez00,Olsen99,Harris01,Dolphin02,
Dolphin13,Yuk07,Walmswell13,Gennaro15}.  This approach is now
routinely applied
\citep[e.g.][]{Weisz08,Weisz11,Weisz13,Weisz14,
Sanna09,Cignoni10,McQuinn10,Hidalgo11,Grocholski12,Monachesi12,
Johnson13,Small13,Bernard15,Lewis15,Williams15} and can yield
accurate, spatially-resolved SFHs, but unfortunately, it can only be
applied to nearby galaxies. For more distant galaxies, galaxy spectra
can be fit using the inversion method to constrain the SFH
(e.g. \citealt{Reichardt01,Panter03,Heavens04,CidFernandes05,Ocvirk06,Tojeiro07,Tojeiro09,Tojeiro13,Koleva09};
see section 4.4 of \citealt{Walcher11}). However, a significant
concern regarding this method for inferring SFHs is that it yields the
smallest number of single-age stellar population templates that fit
the data, which prevents the details of relatively smooth SFHs from
being recovered \citep{Walcher11}.

Photometric spectral energy distribution (SED) modelling based on parametrised SFHs potentially
provides a means to constrain the full SFHs of galaxies
(see \citealt{Walcher11} and \citealt{Conroy13} for recent reviews).
Because broadband photometry requires considerably less integration
time than spectroscopy, the number of galaxies with available
photometry will always be greater than the number with adequate
spectra. Thus, SED modelling potentially provides a means to infer the
SFHs of significantly more galaxies compared with other
methods. Unfortunately, the reliability of SFHs inferred from SED
modelling is unclear (see section 4
of \citealt{Conroy13}). Consequently, most works only attempt to
recover the current SFR and a mass-weighted age. In works that have
attempted to constrain the full SFH, the SFH that corresponds to the
best-fitting SED model is often (explicitly or implicitly) considered
to be the true SFH. However, we shall see
below that this SFH often differs considerably from the true SFH.

Some previous works have presented SED modelling-based methods to
infer parametrised SFHs of galaxies with realistic uncertainties. For
example, \citet{Kauffmann03,Kauffmann03b} combined two stellar
absorption-line indices and broadband photometry to determine
maximum-likelihood estimates of the stellar mass, dust attenuation and
fraction of stars formed in recent bursts for a subset of galaxies
from the
\emph{Sloan Digital Sky Survey} \citep[SDSS][]{york00}. \citet{Mathis06} used the MOPED data compression
algorithm \citep{Heavens00} to extract median-likelihood SFHs from
medium-resolution galaxy spectra from the SDSS.
\citet{Pacifici12} presented a Bayesian method for fitting a combination of photometry and low-to-medium-resolution
spectroscopy to yield the present-day SFR and fraction of stellar mass
formed within the past 2.5 Gyr, among other
parameters. \citet{smethurst15} adopted a simple Bayesian approach to
constrain the SFHs of galaxies assuming a two-parameter SFH model and
fitting to their optical and near-ultraviolet (near-UV)
colours. \citet{Pereira-Santaella15} modified the SED modelling
code \magphys \citep{dacunha08} to yield median-likelihood values for
the time-averaged SFR in four time bins ($0-10$ Myr, $10-100$ Myr,
$0.1 - 1$ Gyr, and $1-10$ Gyr in the past).  However, none of these
works consistently harness the whole range of UV to
far-infrared (far-IR) data to recover the full SFHs of galaxies. Consequently,
a method that provides full SFHs with realistic uncertainties based on
SED modelling of panchromatic photometric data alone remains highly
desirable.

In this work, we present such a method.  Specifically, we demonstrate
how to modify the SED modelling code \magphys\ in order to infer the
SFH of a galaxy by fitting its integrated photometry, expanding the
code's capabilities beyond its original purpose. To validate the
method, we apply it to mock photometry of simulated galaxies generated
by performing three-dimensional (3D) dust radiative transfer on
hydrodynamical simulations of isolated disc galaxies and galaxy
mergers. Because the `true' physical properties of the simulated
galaxies are known and many uncertainties (regarding e.g. the initial
mass function) can be eliminated simply by making identical
assumptions when performing the dust radiative transfer and fitting
the data, this type of controlled experiment is a useful tool for
testing methods of inferring physical properties of galaxies from
observational data
\citep[e.g.][]{lee09,Snyder13,H14,Michalowski14,Torrey15}.
In \citet{HS15}, we used this approach to investigate how
well \magphys could recover various properties of simulated galaxies,
such as the SFR, stellar mass, and dust mass, and to quantify the
effects of physical uncertainties, such as the dust composition. The
success of \magphys at inferring the physical properties of the
simulated galaxies motivated us to undertake the present work.

The remainder of this paper is organised as follows: in \sref{S:methods}, we describe the
SED modelling code \magphys, the proposed method for calculating a median-likelihood SFH, and
the suite of mock SEDs of simulated galaxies used to validate the method. \sref{S:results}
presents the results of applying our method to the simulated galaxies.
In \sref{S:discussion}, we discuss some implications of our results.
\sref{S:conclusions} presents our conclusions.  In this paper we adopt a standard
cosmology with $\Omega_M = 0.3$, $\Omega_\Lambda = 0.7$, and $H_0 =
71$\,km s$^{-1}$ Mpc$^{-1}$.

\section{Methods} \label{S:methods}

\subsection{SED fitting using \magphys}
\label{subsec:magphys}

\begin{figure*}
\centering
\includegraphics[width=1.98\columnwidth,trim=0cm 0cm 0cm 0.7cm,clip]{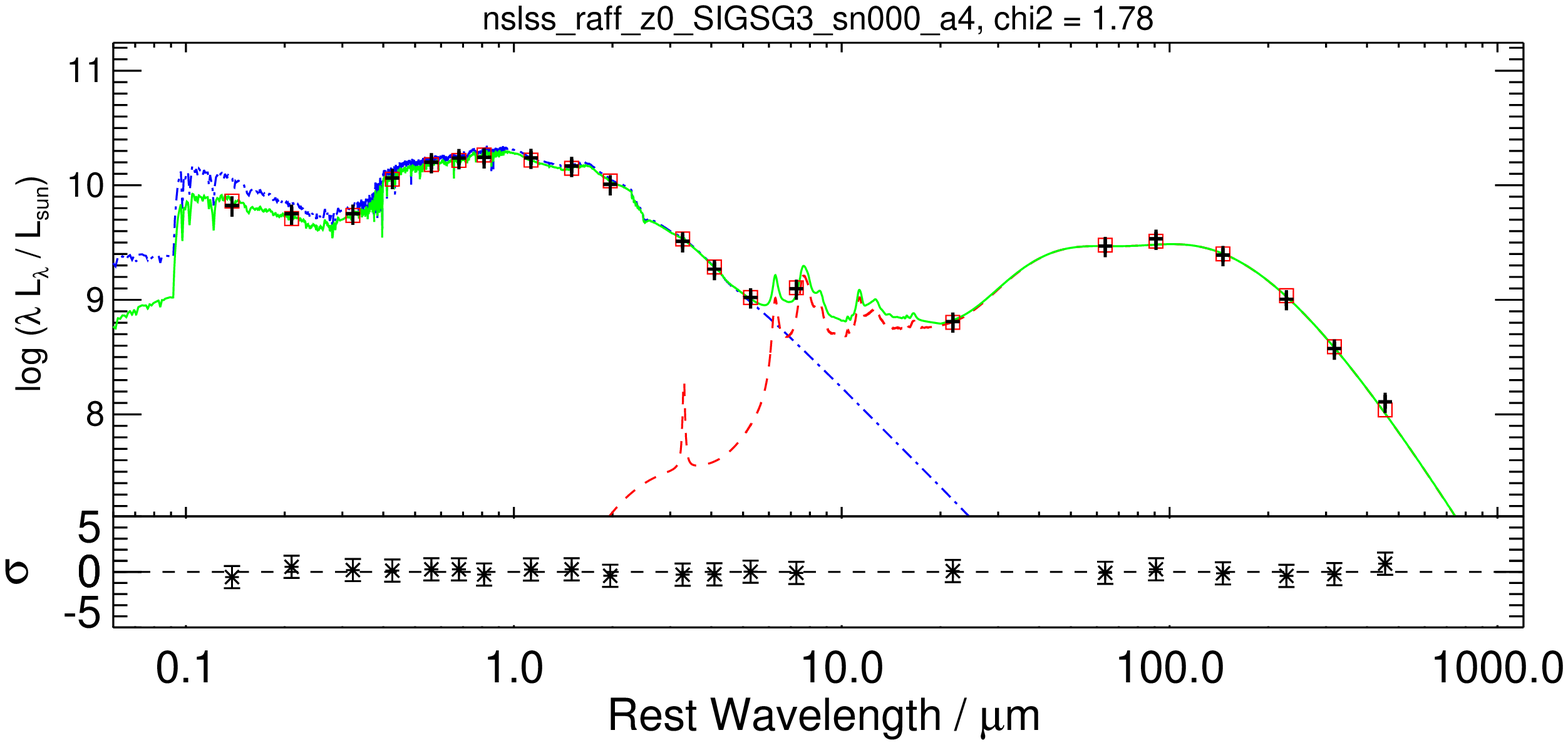}
\caption{An example best-fit UV-mm SED (green solid line). Also
  overlaid are the best-fit unattenuated stellar SED (dot-dashed blue
  line) and the best-fit dust SED component (red dashed line). The
  model photometry associated with the best-fit SED is shown as the
  red squares, while the synthetic photometry derived from the
  simulations is shown as the black error bars (assuming a
  signal-to-noise ratio of 5 in every band). The lower panel shows the
  residuals of the synthetic photometry about the best-fit SED for
  each photometric band; we do not add noise to the photometry, and
  the uncertainties are included solely for the purposes of enabling
  us to use \magphys. }
\label{fig:example}
\end{figure*}

\magphys\footnote{\magphys\ is available from
  \url{www.iap.fr/magphys/}} \citep[][hereafter DC08]{dacunha08} is a
publicly available SED fitting code that assumes an energy balance
criterion to model the
stellar emission of a galaxy consistently with its dust emission. By
assuming that the energy absorbed from the intrinsic starlight by a
two-component dust model \citep[from][with the two components
  corresponding to an ambient diffuse interstellar medium (ISM) and embedded stellar birth
  clouds]{charlot00} is re-radiated in the far-infrared, it is
possible to use the model to not only produce realistic best-fit SEDs
for a wide variety of galaxies with different properties \citep[see
  e.g. ][and references therein]{smith12,HS15} but also to
derive Bayesian probabilistic estimates of their physical parameters
by marginalising over the stellar and dust libraries. 

Here we use the default version of \magphys, which models the emission
from stars using the \citet{chabrier03} initial stellar mass function
along with a library of 50,000 SFHs and stellar spectra taken from the
well-known (unpublished) `CB07' version of the
\citet{bc03} simple stellar population models. The SFHs in the
\magphys\ stellar library consist of two components, a baseline
exponentially declining star formation rate, with bursts randomly
superposed. Approximately half of the SFHs in the library have
experienced a burst in the past 2 Gyr, a feature which is critical for
reliably recovering stellar masses \citep{Michalowski14}.

The dust emission model used in \magphys\ is described in detail in
DC08, but to summarise, each dust SED consists of multiple optically
thin modified blackbody profiles with different normalisations,
temperatures and emissivity indices \citep[see e.g.][for a detailed
  description of modified blackbodies, and an analysis of using them
  to model dust emission in galaxies]{hildebrand83,H12,smith13} describing
dust grains of different sizes, along with a recipe for including
emission from polycyclic aromatic hydrocarbons (PAHs). The primary
adjustable components of the \magphys\ far-IR model are a warm
`birth-cloud' dust component with emissivity index $\beta_{BC} = 1.5$
and temperature between $30 \le T_W^{BC} \le 60$\,K, and a cool
`diffuse ISM' component with $\beta_{ISM} = 2.0$ and
 $15 \le T_C^{ISM} \le 25$\,K, corresponding to the
\citet{charlot00} dust obscuration model applied to the stellar libraries.

\magphys\ combines those stellar and dust-emission libraries
to yield full UV to millimetre (mm) SEDs, which are
then compared with the observed photometry by convolving the
panchromatic models with a set of user-defined filter curves. It then
uses the $\chi^2$ estimator to determine the goodness-of-fit for every
combination of stellar and dust components that satisfies the energy
balance criterion.

In this analysis, we use \magphys\ to fit model SEDs to 21 different
photometric bands, arbitrarily chosen to include data from GALEX at
FUV and NUV wavelengths \citep[e.g.][]{martin05}, the SDSS $ugriz$
bands \citep{york00}, UKIDSS $JHK$ \citep{hewett06}, {\it Spitzer
  Space Telescope} IRAC 3.4 4.5, 5.8 and 8.0\,$\mu$m, MIPS 24 and
70\,$\mu$m \citep[e.g.][]{werner04}, and {\it Herschel Space
  Observatory} data \citep{pilbratt10} at 100, 160, 250, 350 and
500\,$\mu$m. As for our previous work in \citet{HS15}, our goal is to
test \magphys\ under idealised conditions and investigate systematics
rather than effects arising from imperfect observational data
\citep[e.g. the difficulties of cross-identifying {\it Herschel}
  galaxies;][]{smith11}; we therefore do not add any noise to the
input photometry. For the purposes of using \magphys\ for the fitting
however, we arbitrarily assume uncertainties of 20\,per cent in every
photometric band. \fref{fig:example} shows an example best-fit SED
output by \magphys; the synthetic photometry is shown as the black
crosses with error bars while the best-fitting model photometry is
shown by the red squares overlaid on the best-fit emergent SED (in
green). The emergent SED is further decomposed into the best-fit
intrinsic stellar model (blue, dot-dashedd line) and the best-fit IR
template (red dashed line). The lower panel shows the residuals
between the observed photometry and the best-fit model in each band in
$\sigma$ units.

\subsection{Recovering the SFHs and internal validation}
\label{subsec:validation}

In order to extract constraints on galaxy star formation histories
from the public version of \magphys\ we make several modifications to
the code. We first calculate the sum of the relative probabilities,
$P^\prime \equiv \exp(-\chi^2 / 2)$, and the weighted-mean stellar
mass for each SFH in the \magphys\ library, where the averaging is
over every combination of starlight and dust SEDs that satisfies the
energy balance criterion. We then marginalise these 50,000 relative
probabilities over the library of SFHs. This method is analogous to
the way \magphys\ calculates probability distributions for parameters
of interest (e.g. stellar mass), however unlike in the standard
\magphys\ implementation, we retain these data for every galaxy being
studied for the purposes of determining the SFHs in
post-processing\footnote{We are of course able to reproduce the
  \magphys\ stellar mass probability distributions using these data.}.

Since the SFHs in the default \magphys\ library vary in length (due to
the different ages of the continuous component), and since they have
different time resolutions, we linearly interpolate each SFH onto a
common time grid, equally spaced in $\log$ look-back time at intervals
of $\Delta \log T = 0.05$.

Given the marginalised probabilities and the SFHs brought onto a
consistent time resolution, we are able to determine median-likelihood
SFHs by determining the 50$^{\mathrm{th}}$ percentile of the
cumulative distribution of SFR as a function of look-back time. We
also derive uncertainties on the median-likelihood SFH by determining
the 16$^{\mathrm{th}}$ and 84$^{\mathrm{th}}$ percentiles of the
distribution; these values are equivalent to the $\pm 1\sigma$ values
in the limit of Gaussian-distributed uncertainties.

\begin{figure}
\centering
\includegraphics[width=0.98\columnwidth,trim=17.8cm 19cm 8.7cm 0.4cm,clip]{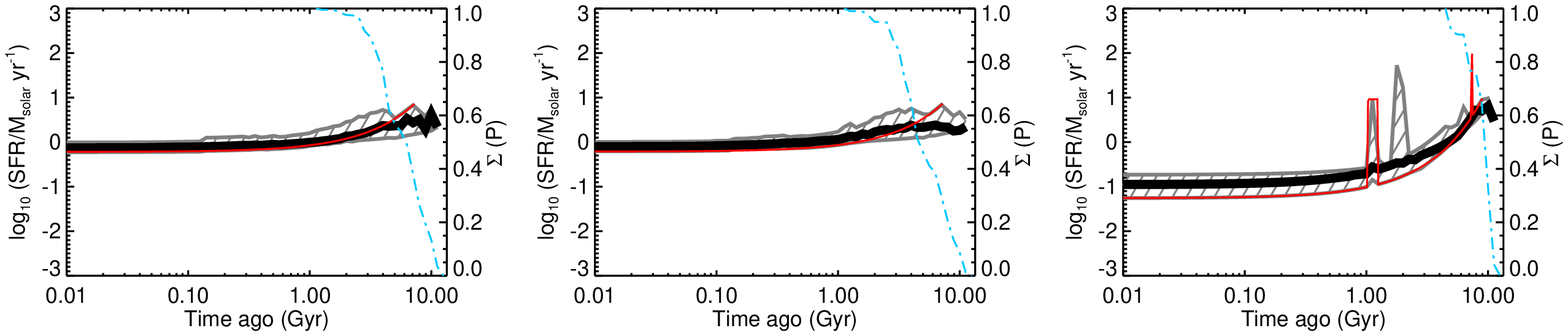}
\caption{Recovering the SFR as a function of look-back time based on
  SED fitting of synthetic photometry for a model $z=0.1$ galaxy taken
  from the default \magphys\ libraries, with assumed 20\,per cent
  photometric uncertainties in every band. The best-fit SFH (which
  also corresponds to the true SFH in this case) is shown as the red
  line, while the median-likelihood SFH is overlaid as the thick black
  line, with the region bounded by the $\pm 1\sigma$ uncertainties
  shaded in grey. The dot-dashed light blue line shows the cumulative
  probability distribution of the SFH as a function of look-back time,
  derived by marginalising over the \magphys\ SFH library, relative to
  the right-hand axis. }
\label{fig:validation}
\end{figure}

An example showing the information that we can derive is shown in
\fref{fig:validation}, with the logarithm of the SFR on the ordinate
and look-back time in Gyr on the abscissa. In this figure, we
internally validate our method by feeding \magphys\ synthetic
photometry derived from one of the SEDs in the default library placed
at $z = 0.1$.  We assume that each photometric datum has an associated
uncertainty of 20\,per cent. The best-fit SFH\footnote{Throughout this
  work, we use the phrase ``best-fit SFH'' to refer to the SFH of the
  SED model that is the best-fit to the {\it photometry}.} (which in
this case corresponds to the true SFH by design, with photometric
$\chi^2 = 0.0$) is shown in as the red line, and we also overlay the
median-likelihood SFH (thick black line) along with the area enclosed
by the $\pm 1\sigma$ uncertainties (grey shaded region). The
median-likelihood SFH is inevitably a worse estimate of the true SFH
than the best fit SFH in this case, given that the exact SFH of this
galaxy is in the library; however, the true SFH is always within
1$\sigma$ of the median-likelihood values (including during the strong
burst of star formation around 1\,Gyr ago). We attribute the fact that
the SFH uncertainties are not centred around the best-fit\slash true
SFH to the prior distribution of SFHs in the \magphys\ library (we
will return to the topic of the SFH priors in what follows, however we
note that the median-likelihood SFHs we recover here and elsewhere are
considerably different from the median SFH of the \magphys\ library,
indicating that useful SFH constraints are being obtained from the
photometry). We also calculate the total SFH probability (i.e. the sum
of the probability in all SFHs defined at any given look-back time) as
a function of look-back time; this is overlaid as the light-blue
dot-dashed line in \fref{fig:validation}, with values indicated by the
right-hand axis.

To quantify how well we are able to recover the SFHs of individual
simulated galaxies, we define two parameters, \chisqsfh\ and
\dM, as follows:

\begin{equation}
\chi^2_{\rm{SFH}} = \frac{1}{N} \sum_t \frac{\left({\rm SFR}_{{\rm model}}(t) - {\rm SFR}_{{\rm true}}(t)\right)^2}{\sigma_{\rm SFR}(t)^2},
\label{eq:chi2_sfr}
\end{equation}

\noindent and, 

\begin{equation}
\frac{\Delta M}{M} = \frac{\sum_t | {\rm SFR}_{\rm model}(t) - {\rm SFR}_{\rm true}(t) |}{\sum_t {\rm SFR}_{\rm true}(t)},
  \label{eq:deltaM}
\end{equation}

\noindent where ${\rm SFR}_{\rm model}(t)$ represents either the SFH
associated with the best-fitting SED template or the median-likelihood
SFH, ${\rm SFR}_{\rm true}(t)$ is the known SFH of the simulation,
$\sigma_{\rm SFR}(t)$ is the uncertainty on the median-likelihood SFH
as a function of look-back time, and the summations are over the N
bins in look-back time for which both the true SFH and the SFH being
compared with are defined. \chisqsfh\ is thus a measure of how well
any given SFH (e.g. the best-fit or median-likelihood that we recover)
tallies with the true SFH that we know from the simulations (after
accounting for the uncertainties), though we emphasize that we do not
use use this parameter in any SED fitting.\footnote{It is also worth
  noting that \chisqsfh\ intrinsically favours shorter SFHs (as they
  have a lower number of measurements); given that median-likelihood
  SFHs are defined at all times where the \magphys\ library contains
  at least one SFH (i.e. over the whole Hubble time), one might expect
  that \chisqsfh\ would favour the best-fit SFHs. We do not attempt to
  account for this effect in what follows (e.g. by introducing a
  `reduced' \chisqsfh).} \dM\ quantifies the integrated absolute
difference between the model and known SFH as a fraction of the total
mass of formed stars. Note that because the absolute difference is
used, \dM\ can be large even if the stellar mass is accurately
recovered (because for the mass, time periods in which the true SFR is
overestimated can be compensated for by time periods in which it is
underestimated). In what follows, we will use these two parameters to
inform our discussion of the results of using \magphys\ to estimate
the SFHs of simulated galaxies.

\subsection{Simulations used for validation}
\label{subsection:simulations}

To validate the method, we apply it to mock SEDs generated from
hydrodynamical simulations of isolated disc galaxies and binary galaxy
mergers (see \citealt{HS15} for a detailed discussion of the merits of
this type of external validation). Because the SFHs of the simulated
galaxies are known, this approach enables us to test how well our method can successfully
recover the true SFH from photometry alone. We do not add noise to the
mock photometry; consequently, we test whether physical limitations
prevent us from recovering the SFH even when we have perfect
(i.e. noiseless) data.

We utilise a subset of the mock SEDs from the suite of simulations
first presented in \citet{Lanz14}.\footnote{The SEDs are publicly
  available at \url{http://dx.doi.org/10.7910/DVN/SIGS_SIMS_I}.}  The
full dataset contains SEDs for four isolated disc galaxy simulations
with stellar masses that range from $6 \times 10^8$ to $4 \times
10^{10} \msun$ and binary mergers of all possible combinations of
progenitors (i.e. 10 mergers) for a single generic orbit. The merger
mass ratios range from 1:1 to 1:69. The progenitor galaxies were
designed to have properties (e.g. gas fractions) that are typical of
galaxies in the local Universe; see \citet{Cox08} for details. The
progenitor discs are referred to as M0, M1, M2, and M3, in order of
increasing stellar mass. The mergers are referred to using the labels
of the two progenitors followed by an `e' (because the `e' orbit of
\citealt{Cox08} was used), e.g. M3M2e. In this work, we present
results from the isolated disc simulations (M0, M1, M2, and M3) and
the M3M2e merger simulation.

\begin{figure*}
\centering
\subfigure{\includegraphics[width=1.8\columnwidth,trim=8.8cm 12.5cm 8.8cm 0.0cm,clip]{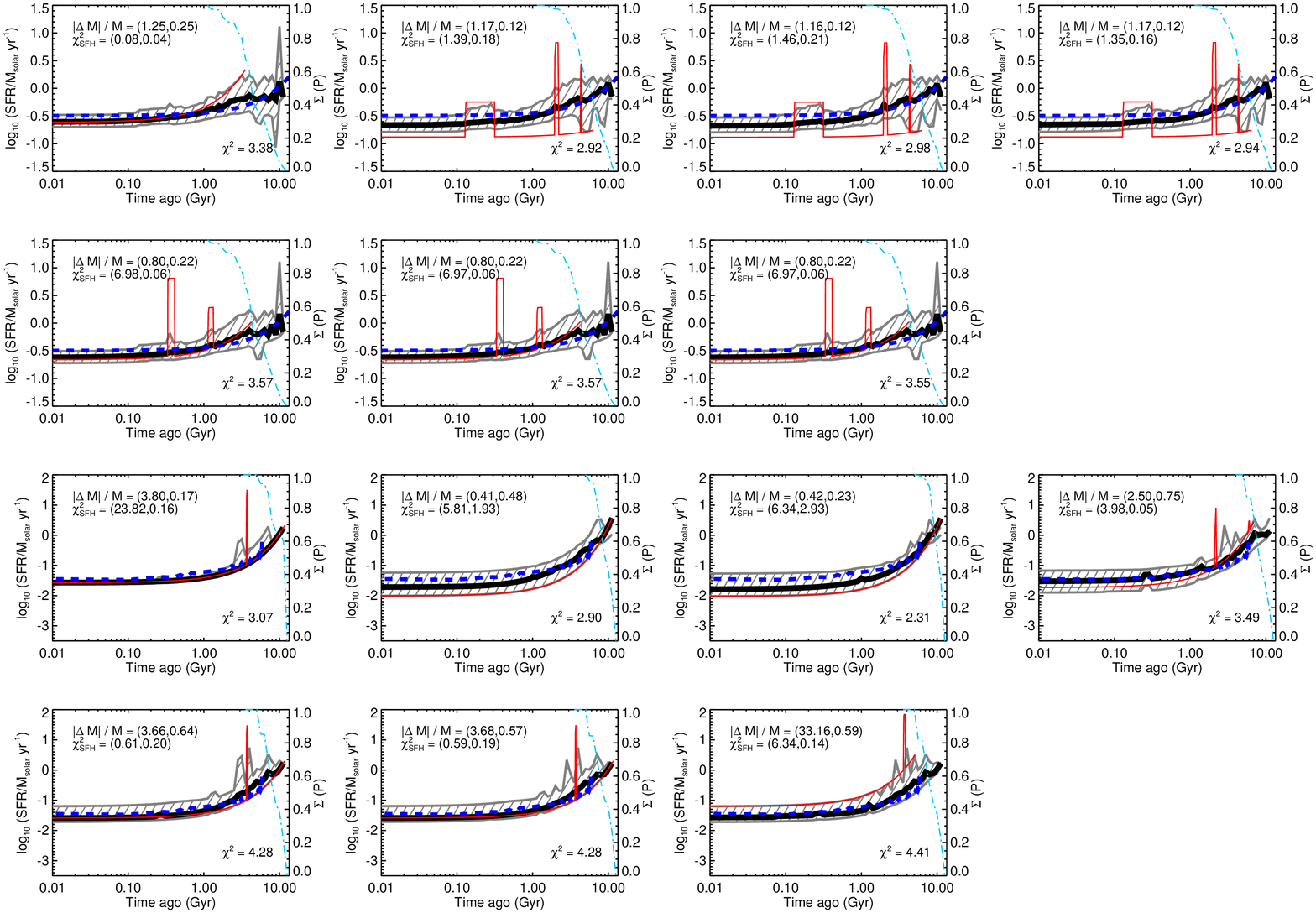}}
\caption{Constraints on the SFH of the first snapshot of the M2
  simulation, where each panel shows the results for one of the
  individual viewing angles, modelled using \magphys. The true SFH is
  shown by the dashed blue line, the best-fit estimate produced by
  \magphys\ is shown as the red solid line, and the median-likelihood
  SFH is shown as the thick black line. The region enclosing the
  16$^{\mathrm{th}}$ and 84$^{\mathrm{th}}$ percentiles of the SFH PDF
  at each look-back time is shown by the grey shaded region. The
  legend in the upper left of each panel details the values of $\Delta
  M \slash M$ and $\chi^2_{\rm SFH}$ for the best-fit and
  median-likelihood SFHs, respectively, while the lower-right legend
  shows the best-fit value of $\chi^2$ for each model. The light-blue
  dot-dashed line shows the SFH cumulative frequency distribution
  derived by marginalising over the \magphys\ SFH library, relative to
  the right-hand axis. While the SFH of the model with the best fit to
  the photometry shows variation with viewing angle and spurious
  bursts not present in the true SFH, the median-likelihood SFH
  estimate is better-behaved.}
\label{fig:sfh_recovery_first}
\end{figure*}

We will now briefly summarise the details of the hydrodynamical
simulations and mock SED generation, but we refer the reader to
\citet{Lanz14}, \citet{HS15} and references therein for full
details. First, idealised galaxies composed of a dark matter halo,
gaseous and stellar discs, stellar bulge, and supermassive black hole
were created following the procedure of
\citet{Springel05feedback}. Then, the dynamical evolution of each of
the isolated discs and mergers was simulated using a modified version
of the $N$-body/smoothed-particle hydrodynamics
(SPH)\footnote{Recently, it has been demonstrated that simulations
  performed using the traditional density-entropy formulation of SPH,
  which is employed in the version of \gadgettwo used for these
  simulations, suffers from significant numerical inaccuracies that
  can qualitatively affect the results of galaxy formation simulations
  \citep[e.g.][]{Agertz:2007,Springel:2010arepo,Bauer:2012}. However,
  the type of simulations used for this work are relatively
  insensitive to these inaccuracies \citep{H14arepo}, so the use of
  traditional SPH should not be cause for concern.} code \gadgettwo
\citep{Springel05gadget}. The simulations directly include the effects
of gravity, hydrodynamics, and radiative heating and cooling. The SFRs
associated with individual gas particles are calculated according to a
volume-density-dependent Kennicutt--Schmidt relation
\citep{Schmidt:1959,Kennicutt:1998} with a low-density threshold. Star
particles are stochastically spawned from gas particles, where the
probability that a given gas particle spawns a star particle is
proportional to its SFR. Supernova feedback is modeled using the
two-phase interstellar medium model of \citet{Springel03}. Metal
enrichment is treated by evolving each gas particle as a closed box.
Black hole accretion and active galactic nucleus (AGN) feedback are
included as described in \citet{Springel05feedback}.

At various times throughout the simulations (every 10 or 100 Myr;
times at which the SFR varies rapidly were sampled more frequently),
`snapshots' of the physical state of the simulation were saved. Then,
in post-processing, three-dimensional (3D) dust radiative transfer was
performed using the Monte Carlo radiative transfer code \sunrise
\citep{Jonsson06,Jonsson10}. This process proceeds as follows.  First,
the sources of radiation are specified: the star particles are
assigned \emph{Starburst99} \citep{Leitherer:1999} single-age stellar
population SEDs according to their ages and metallicities.  The
progenitor galaxies include stellar discs and bulges, and these star
particles must be assigned ages and metallicities. The stellar disc is
assumed to have formed with an exponentially declining SFH, whereas
the bulge is assumed to have formed via an instantaneous burst. The
metallicities of the stars that exist at the start of the simulations
and the initial gas metallicity are specified via a profile that
decreases exponentially with distance from the galaxy center. Both the
SFHs for the stellar disc and bulge and the metallicity gradients have
been constrained by comparisons with observations of local galaxies;
see \citet{rocha08} for details. The star particles formed in the
simulations have ages and metallicities that are determined
self-consistently. We note that the resulting SEDs are rather
insensitive to the assumed SFH for the stellar disc and bulge and
metallicity gradient, especially after the first few hundred Myr
\citep[e.g.][]{H11}.  The AGN particles are assigned
luminosity-dependent template SEDs from \citet{Hopkins:2007}, which
are based on observations of unreddened quasars. Subsequently, the
dust distribution is calculated by projecting the metal content of the
gas particles onto a 3D octree grid, assuming a dust-to-metals ratio
of 0.4 \citep{Dwek:1998,James:2002}.

With the source positions, source SEDs and dust distribution in hand,
radiative transfer is performed to calculate the effects of dust
absorption and scattering. The thermal-equilibrium temperatures of
dust grains, which depend on the local radiation field and the
wavelength-dependent grain opacity, are calculated. Subsequently,
radiation transfer of the resulting IR emission is performed. To
account for dust self-absorption, the dust temperature calculation and
IR radiation transfer are iterated until the temperatures converge.
The calculation results in spatially resolved UV--mm SEDs of the
galaxies viewed from multiple viewing angles (seven in our case). We
sum the SEDs of all individual pixels to obtained galaxy-integrated
SEDs and then convolve these with the appropriate filter response
curves to obtain broadband photometry.

\section{Results} \label{S:results}

In this section we will determine how well \magphys\ can recover SFHs
for two classes of simulations in which the answer is known, isolated
disks and galaxy mergers. Both classes are of important diagnostic
value: the isolated disk SFHs should be reasonably well-described by
the simple exponentially-decaying or ``$\tau$ model'' SFH
parametrisations in \magphys, whilst we expect that the galaxy mergers
have more complex and ``bursty'' SFHs (and as discussed in
\sref{subsec:magphys}, bursts are randomly superposed on
the \magphys\ SFHs).

\subsection{Isolated disc SFHs}

\fref{fig:sfh_recovery_first} compares the best-fit and
median-likelihood SFHs for four of the seven viewing angles from the
first snapshot of the M2 simulation.  In each panel, the true SFH
(which is independent of viewing angle, of course) is shown as the
dashed blue line; because this is the first snapshot, the SFH is that
assumed for the stars that exist at the start of the simulation i.e.
an exponentially declining SFH for the disc stars and an instantaneous
burst for the bulge (see \sref{subsection:simulations} for
details). The best-fit SFH derived using \magphys\ is shown by the red
line, and the median-likelihood SFH is shown as the thick black line,
with associated uncertainties indicated by the grey shaded
region. The \dM\ and \chisqsfh\ values for each viewing angle are
shown in the upper-left legend in each panel, while the lower-right
legends show the best-fit $\chi^2$. The dot-dashed light blue line in
each panel shows the SFH cumulative frequency distribution of the
model galaxy (relative to the right-hand axis), which is derived by
marginalising over the \magphys\ library of SFHs for each viewing
angle.

It is immediately apparent from \fref{fig:sfh_recovery_first} that the
best-fit SFH (red line) can vary depending on the viewing angle, while
the median-likelihood SFH (thick black line) derived by marginalising
over the SFHs in the default \magphys\ library is rather more
consistent. Furthermore, the best-fit SFH often falls outside the grey
shaded region (which represents the range of $\pm 1\sigma$ on the SFH
at each snapshot); in 3\slash 7 cases it is systematically offset,
while in a further 3 cases the best-fit SFH indicates the presence of
starbursts which are not present in the true SFH.  In stark contrast,
the median-likelihood SFH is in good agreement with the true SFH at
all values, once the uncertainties are taken into account. This is
true even at large look-back times, where the median-likelihood SFH
estimates and the associated uncertainties become angular and noisy;
we attribute this effect partly to the difficulty of distinguishing
between stellar populations older than $\sim 1$\,Gyr, and partially
due to the small number of SFHs in the \magphys\ library that give
acceptable fits to the synthetic photometry with sufficiently large
ages. This effect is underlined by the plunge in the SFH cumulative
probability distribution (dot-dashed light-blue lines) in each panel
of
\fref{fig:sfh_recovery_first}. The \dM\ values (representing the
fractional mass discrepancy) are lower for the median-likelihood SFHs
in each case; the best-fit SFHs containing bursts are also strongly
disfavoured by the \chisqsfh\ values.

\begin{figure}
\centering 
\subfigure{\includegraphics[width=\columnwidth,trim=1cm 0cm 0cm 0.0cm,clip]{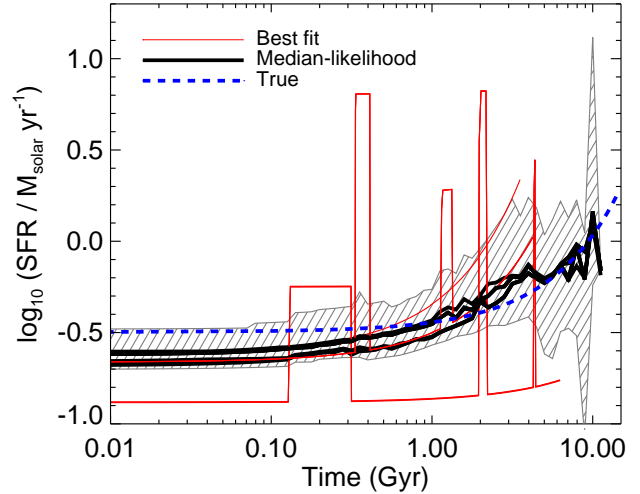}}
\caption{SFHs recovered for the seven different viewing angles to the
  first snapshot of the M2 isolated disk simulation shown
  in \fref{fig:sfh_recovery_first}. The true SFH is shown as the blue
  dashed line, while the best-fitting SFHs for each of the seven
  viewing angles are shown as the red lines. The median-likelihood
  SFHs for each viewing angle are shown as the thick black lines,
  while the shaded grey region shows the average uncertainty
  associated with the median-likelihood SFHs. The contrast between the
  SFH of the best-fit model and the median-likelihood SFH (in terms of
  both reliability and fidelity) is clear. }
\label{fig:angle_comparisons}
\end{figure}

To ease comparison, and show all seven viewing angles, we overlay the
individual best-fit and median-likelihood SFHs for each angle of the
first snapshot in the M2 simulation with one another in
\fref{fig:angle_comparisons}. It is immediately apparent that
the best-fit SFHs (in red) are much less consistent between angles and
show worse agreement with the true SFH (dashed blue line) than the
median-likelihood SFHs (thick black lines). To test whether this
behaviour is due to the choice of prior, we re-run our internal
validation discussed in section \ref{subsec:validation}, excluding the
true SFH from the model library. We find that under these conditions
our internal validation returns a similar disagreement between the
best-fit and true SFHs, suggesting that the best-fit SFH is unreliable
even with a realistic prior on the SFHs. We speculate that this
behaviour may be due to parameter degeneracies even in panchromatic
broad-band galaxy SEDs.

\begin{figure*}
\centering \includegraphics[width=1.8\columnwidth,trim=8.8cm 0cm 8.8cm
  12.5cm,clip]{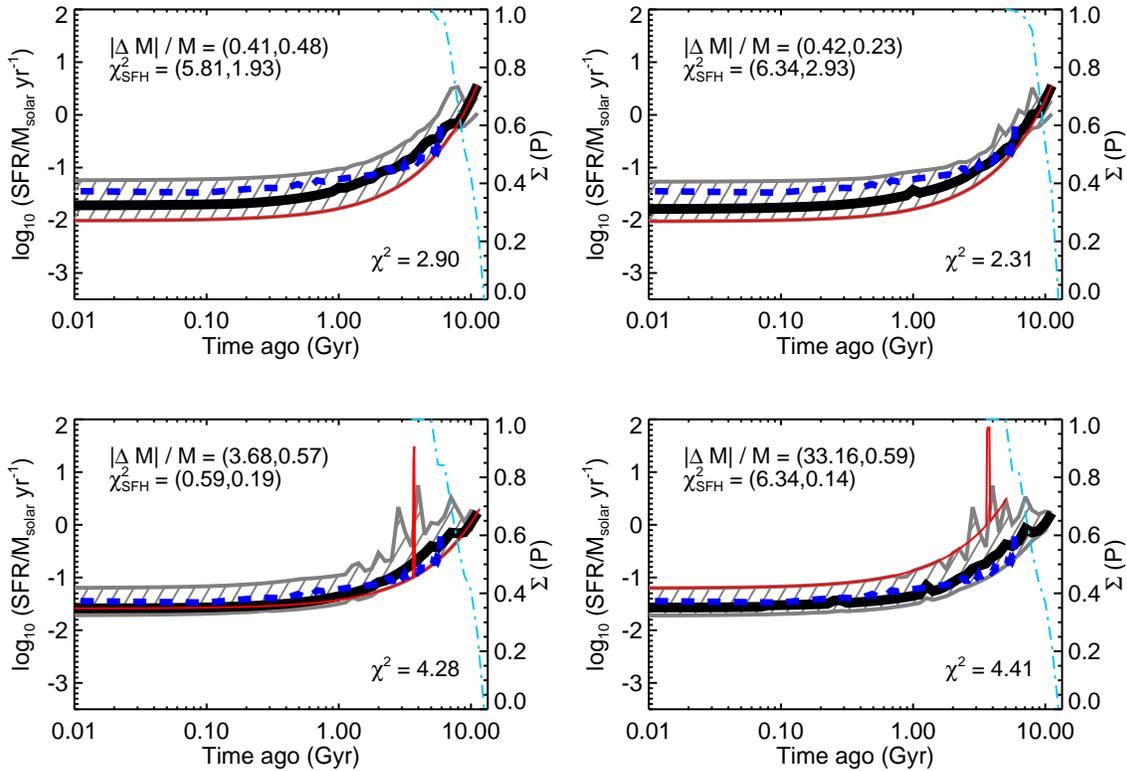}
\caption{SFHs recovered for the final snapshot of the M2 simulation;
  the true SFH compiled from the individual simulation snapshots is
  shown by the dashed blue line, while the best-fit SFH from \magphys\
  is overlaid with a red solid line. The thick black line represents
  the median-likelihood SFH, while the grey region shows the range of
  $\pm 1\sigma$ about the median-likelihood SFH as a function of
  look-back time. The legends are as
  in \fref{fig:sfh_recovery_first}. The median-likelihood SFH is once
  again preferred over the SFH of the model that is the best-fit to
  the photometry.}
\label{fig:sfh_recovery_last}
\end{figure*}

In \fref{fig:sfh_recovery_last} we compare four of the SFHs
constructed using the modified \magphys\ with the true SFH for the
last snapshot of the simulation.
This presents a useful test: because stars (more precisely, star particles
that are analogous to star clusters) have formed throughout the
simulation according to the assumed volume-density-dependent
Kennicutt-Schmidt relation (see \sref{subsection:simulations} for
details), these SFHs do not have a simple, generic analytic
form, as can be seen in \fref{fig:sfh_recovery_last}.

\begin{figure}
\centering \includegraphics[width=\columnwidth,trim=1cm 0cm 0cm 0.0cm,clip]{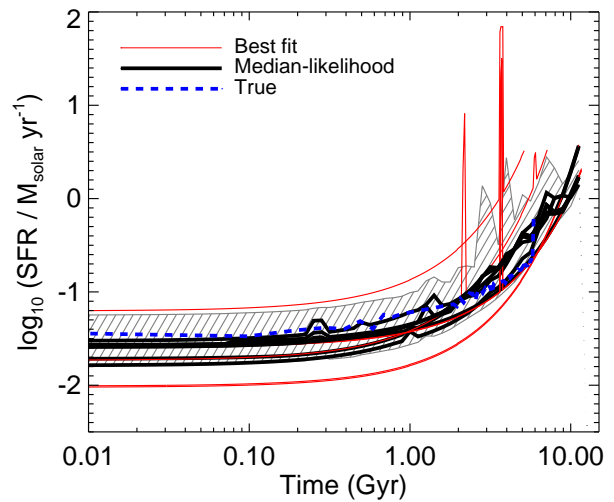}
\caption{SFHs recovered for the seven different viewing angles to the
  last snapshot of the M2 isolated disk simulation shown
  in \fref{fig:sfh_recovery_last}. The true SFH is shown as the dashed
  blue line, while the best-fitting SFHs for each viewing angle are
  shown by the red lines. The median-likelihood SFHs are shown as
  thick black lines, while the shaded grey region shows the average
  uncertainty associated with the median-likelihood SFHs. The
  median-likelihood SFH is more reliable and a better approximation of
  the true SFH than the SFH of the best-fit model.}
\label{fig:angle_comparisons_last}
\end{figure}

Once more, the \chisqsfh\ and \dM\ values point to greater
fidelity in the median-likelihood SFHs rather than the best-fit
values, which also show greater variation with viewing angle and the
presence of bursts which do not exist in the true SFH.
This variation is more apparent in
\fref{fig:angle_comparisons_last}, in which we directly overlay
the best-fit and median-likelihood SFHs for each of the seven viewing
angles. The colour scheme is as in
Figs. \ref{fig:sfh_recovery_first}, \ref{fig:angle_comparisons} \&\ \ref{fig:sfh_recovery_last}. The
true SFH constructed from the individual snapshots of the M2
simulation is arguably a better test of \magphys\ than the previous
tests, since it should be more realistic for an evolving disk
galaxy. That there is such good agreement between the
median-likelihood SFH derived using \magphys\ and the true SFH offers
considerable encouragement for using \magphys\ in this way.

\subsection{Galaxy merger SFHs}

We now turn our attention to the M3M2e simulation, corresponding to a
major galaxy merger with a mass ratio of 2.3:1. We examine, in
particular, two of the time snapshots after the individual components
have coalesced, at which point \magphys\ is able to produce an
acceptable fit to the model photometry. These snapshots are of
particular interest, since the true SFHs drawn from the simulations
are dominated by a recent, extended, and merger-induced burst of star
formation and are considerably more complicated than the simple
``exponentially declining + burst'' SFHs assumed in the version
of \magphys\ used here.  As a result, they present an excellent test
of how well \magphys\ can perform under this particularly challenging
scenario using the standard libraries.

\begin{figure*}
\centering 
\subfigure{\includegraphics[width=\columnwidth,trim=0cm 0cm 0cm 0.0cm,clip]{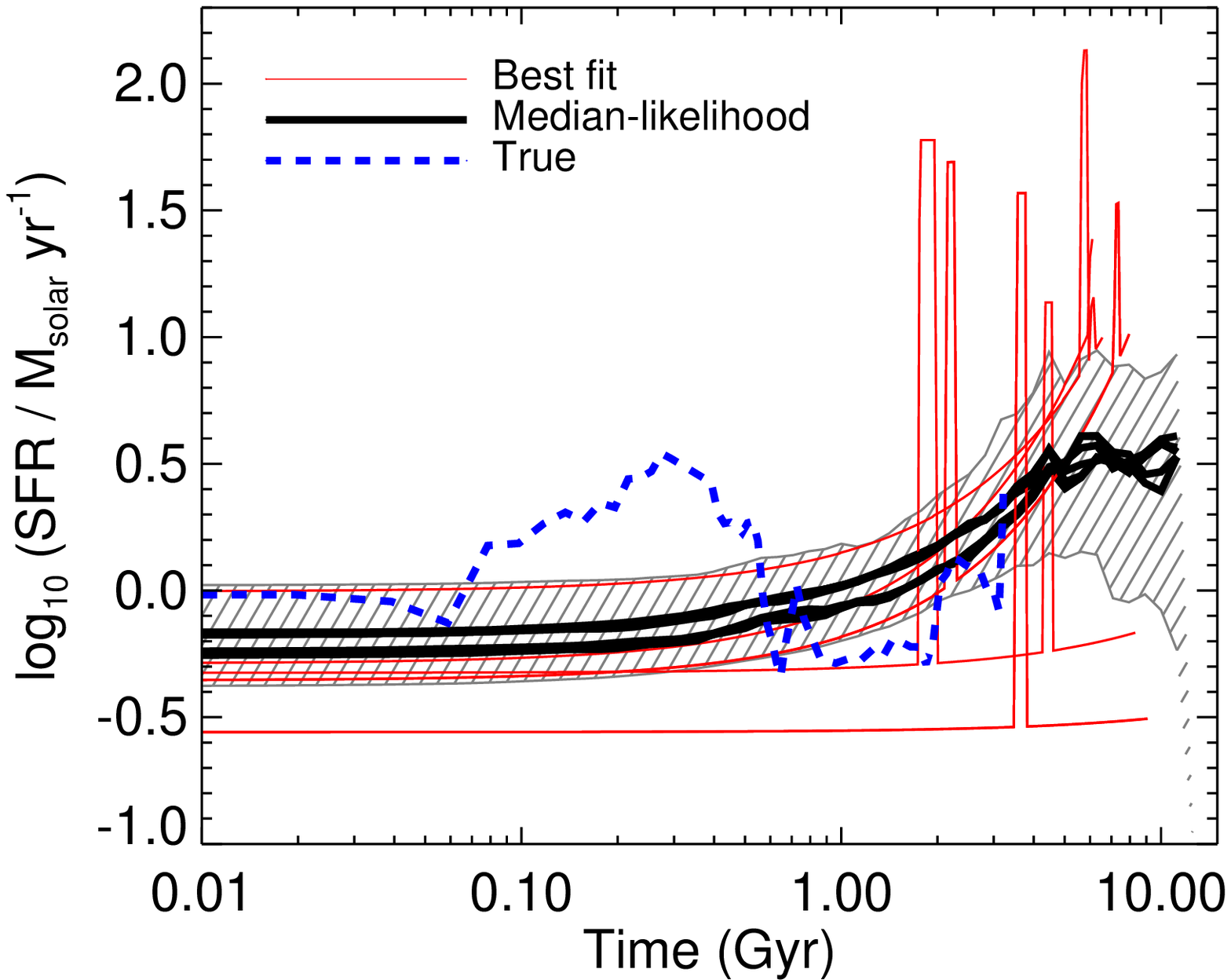}}
\subfigure{\includegraphics[width=\columnwidth,trim=0cm 0cm 0cm 0.0cm,clip]{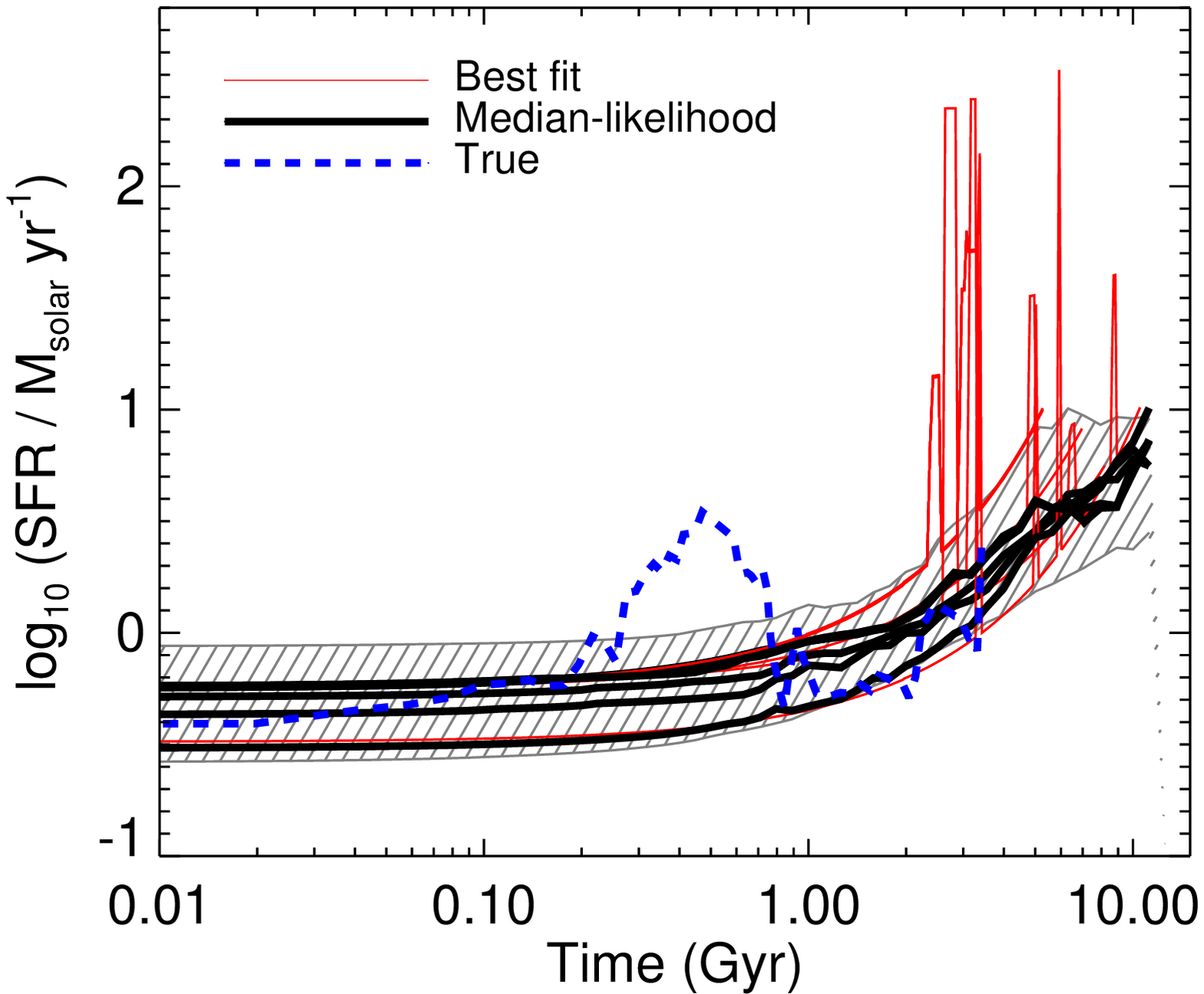}}
\caption{SFHs recovered for the seven different viewing angles for two
  snapshots taken from the M3M2e merger simulation. The true SFH is
  again shown as the dashed blue line, while the best-fitting SFH for
  each viewing angle is shown by the red lines. The median-likelihood
  SFHs are shown as thick black lines, while the shaded grey region
  shows the average uncertainty associated with the median-likelihood
  SFHs. The {\bf left} plot shows the SFH recovered from the 60th
  snapshot, corresponding to $\sim 0.3$\,Gyr after the time of the
  peak in the merger-induced starburst, while the {\bf right} plot
  shows the SFH recovered for the 70th snapshot, $\sim 0.5$\,Gyr after
  the peak SFR. Neither the best-fit or median-likelihood SFH is able
  to recover the main burst of merger-induced star formation.}
\label{fig:merger_comparisons}
\end{figure*}

\fref{fig:merger_comparisons} presents the results for snapshots taken
around 0.3 and 0.5\,Gyr after the peak of the merger-induced starburst
in the left- and right-hand panels, respectively. The starburst can be
clearly seen in the true SFHs (blue dashed lines), though the SFHs
reconstructed from \magphys\ (whether they are best-fit or
median-likelihood) are clearly incorrect, despite the $\chi^2$ values
indicating a good fit to the photometry in both cases \citep[and
despite][having demonstrated that it is still possible to derive
reasonable e.g. SFRs and stellar masses under similar
conditions]{HS15}. Once more the \magphys\ library SFHs associated
with the best-fit to the photometry are littered with spurious bursts
of star formation, thus highlighting the difficulty in interpreting
burst-related properties.  We shall return to these points
in \sref{S:discussion}.

\section{Discussion} \label{S:discussion}

\subsection{Median-likelihood vs. best-fitting SFHs}

In \sref{S:results}, we noted that the median-likelihood SFH estimates
are more consistent with viewing angle and agree better with the true
SFH than the best-fit SFH that we derive. Above, we showed a few
examples to demonstrate our method and highlight the merits of the
median-likelihood SFHs. To assuage any concerns that we have only
shown the best examples and present a more complete analysis, we now
compare the two possibilities quantitatively by using equations
\ref{eq:chi2_sfr} and \ref{eq:deltaM} to calculate \chisqsfh\ and
\dM\ for each of the seven viewing angles to the first and last
snapshots of the isolated disk simulations, for which \magphys\ recovers a
good fit. We have not included the merger simulations because the SFHs
are generally not well recovered owing to the merger SFHs differing
drastically from those assumed in the standard \magphys\ library.
\chisqsfh\ and \dM\ tell us how well the SFH is recovered by our
modified version of \magphys, though we emphasize again that these
parameters are not included in the SED fitting itself (since it can be
rather difficult to know the true SFH for a real galaxy a
  priori).

\begin{figure*}
\centering 
\subfigure{\includegraphics[width=\columnwidth,trim=0cm 0cm 0cm 0.0cm,clip]{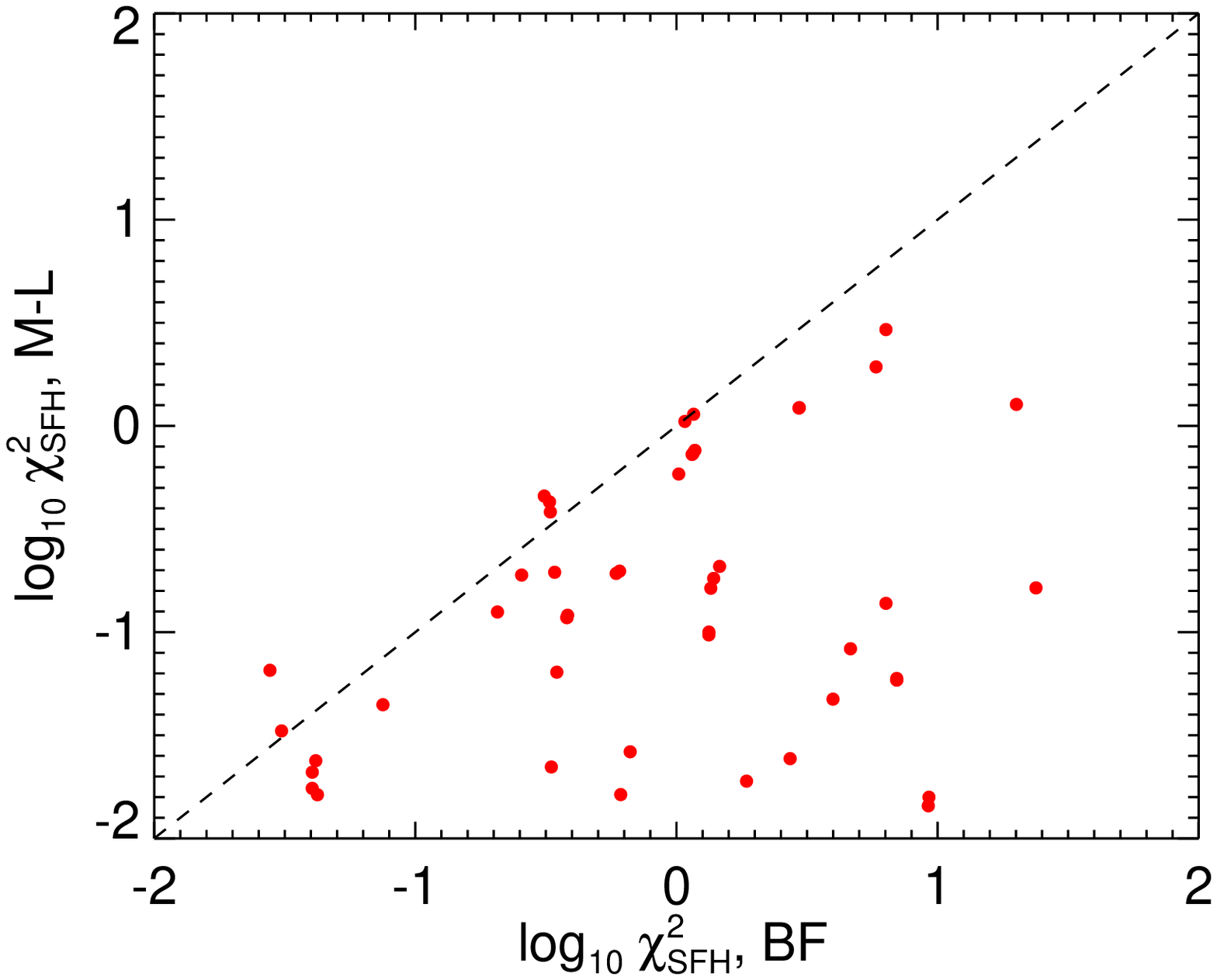}}
\subfigure{\includegraphics[width=\columnwidth,trim=0cm 0cm 0cm 0.0cm,clip]{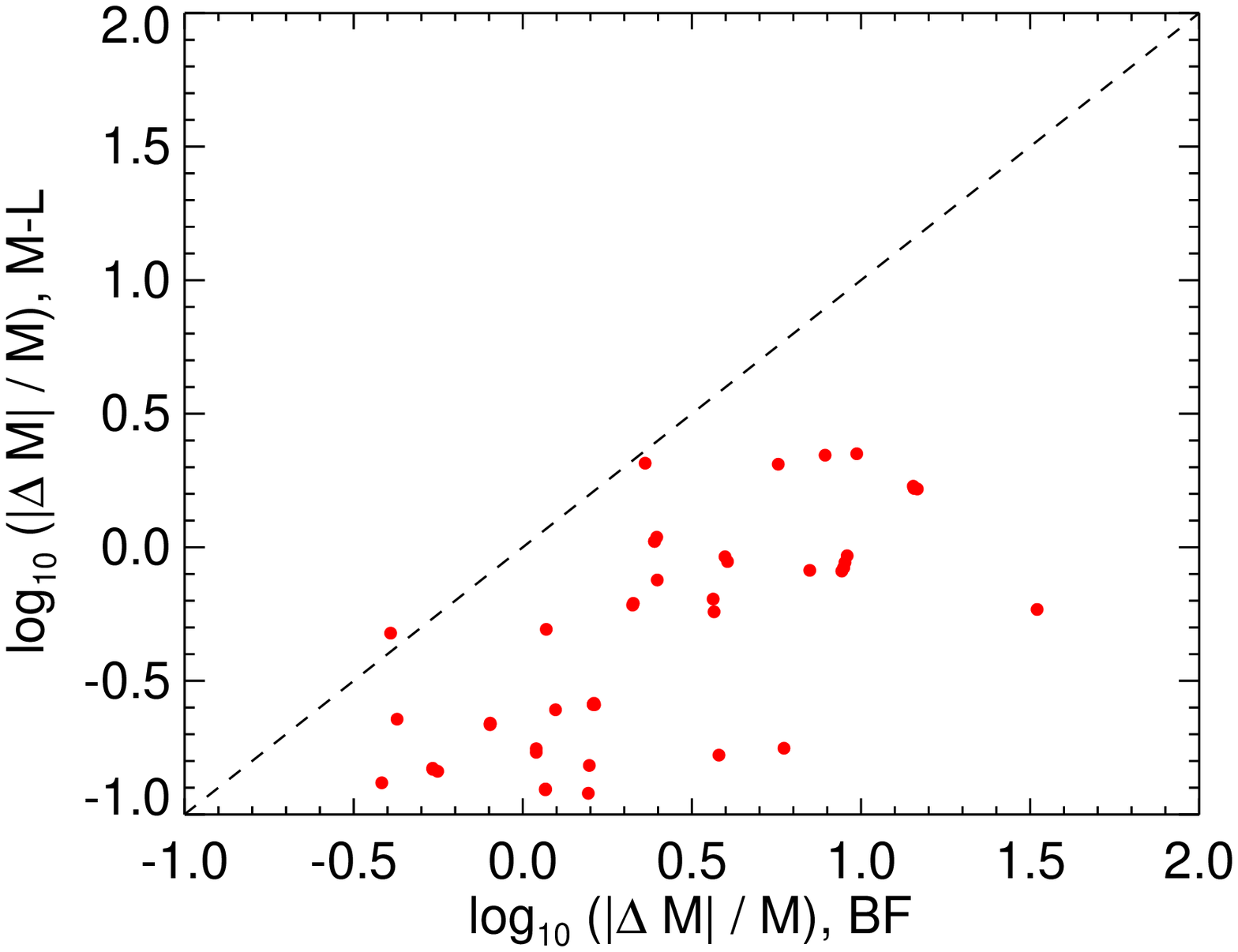}}
\caption{Comparison between the values of \chisqsfh\ (left panel) and
  \dM\ (right panel) for the first and last snapshots of the isolated
  disk simulations, calculated for the median-likelihood (``M-L'', on
  the $y$-axis) and best-fit (``BF'', on the $x$-axis) SFHs derived
  using our modified version of \magphys. The dashed line indicates
  parity, i.e. no preference for either type of SFH. In the right-hand
  panel $\log_{10}($\dM$)=0.0$ implies an absolute integrated mass
  discrepancy between the true and model SFH that is equal to the
  present-day stellar mass. Both the \chisqsfh\ and \dM\ comparisons
  indicate that the median-likelihood SFHs tend to better represent
  the true SFHs.}
\label{fig:parameters}
\end{figure*}

The left-hand panel of \fref{fig:parameters} shows a comparison of the
values of \chisqsfh\ returned for the best-fit (``BF'', shown on the
$x$-axis) and median-likelihood (``M-L'', on the $y$-axis) SFHs
derived by our modified version of \magphys. The fact that the vast
majority of the data points lie below the dotted line (indicative of
parity) highlights that our analysis strongly favours the
median-likelihood SFHs over the individual SFH that provides the best
fit to the photometry. The right-hand panel shows the same preference
for the median-likelihood SFHs in terms of \dM, indicating the best-fit SFHs
show a larger absolute stellar mass discrepancy than the
median-likelihood SFHs (this is expected given
their poorer \chisqsfh). We note that it is possible to have a large
\dM\ and still recover a reasonable stellar mass estimate, since
\dM\ is extremely punitive. This is because \dM\ accounts for the {\it
  time} at which the stars are formed, whereas the stellar mass
  can be recovered accurately if times at which the SFR is overestimated
  are compensated for by times at which it is underestimated.
  $\log_{10} ($\dM$)=0.0$ implies
an absolute integrated mass discrepancy between the true and model SFH
that is equal to the present-day stellar mass. In \citet{smith13} we noted a slight preference for
median-likelihood parameter estimates (e.g. dust luminosity, or
isothermal dust temperature) due to their slightly lower bias relative
to the best fit parameters; here the preference for the median
likelihood SFHs is rather stronger.

Perhaps the most obviously unsatisfactory features of the best-fit
SFHs are the inconsistency with viewing angle and the unreliable
behaviour of the bursts. In the former case, the inconsistency with
viewing angle of the best-fit SFHs is of particular concern for real
observations, where the line-of-sight to any extragalactic object is
fixed. Regarding the latter issue, the best-fit SFHs often include
spurious bursts for the isolated disk simulations (where they should
not be present). For the the merger simulations (where they
should be present), bursts appear at the wrong point in the
SFH. The median-likelihood SFHs can mitigate the viewing angle
dependence and show no evidence for spurious bursts of star formation
for the isolated disk simulations. However, they are also unable to
approximate the complex SFHs of the merger simulations, likely because
the simple analytic form for the SFHs contained in the standard
\magphys\ library is too restrictive \citep[it contains bursts with a
  constant, elevated SFR that last between 30 and
  300\,Myr;][]{dacunha08}; we will discuss this issue in detail below.

That the best-fit SFHs appear so unreliable in comparison to the
median-likelihood values is perhaps not surprising: if the true SFHs
are not present in the \magphys\ prior, then it is only by
marginalising over the SFH library that we could hope to recover
something approaching the truth. This is also the case in the real
Universe: we cannot reasonably expect synthetic libraries to contain
every possible galaxy SFH (even if they did have an analytic form).
This provides further motivation for adopting our statistical approach
to deriving realistic galaxy SFHs from photometry.

\subsection{The need for more complex SFHs}

That our method was unable to recover SFHs of the major mergers is
perhaps not surprising, given the complex form of the merger
simulations' SFHs. We are unable to approximate such SFHs even by
marginalising over the entire \magphys\ library, though the merger
simulation SFHs are by no means the most complex or extreme that exist
in the real Universe \citep[or even the latest simulations
  e.g.][]{hopkins14}. We suggest that it would be extremely desirable
to include more complex SFHs in the \magphys\ libraries if we wish to
use it to study the individual star formation histories of galaxies in
detail based on photometry alone. Though they represent a succinct and
physically-motivated means of describing rudimentary composite stellar
populations, the shortcomings of the so-called ``$\tau$ models'' are
clear \citep[see e.g.][]{lee09,lee10,Conroy13,Simha14} and recent
studies have noted a preference for delayed $\tau$ models
\citep[consisting of a linear rise preceding the exponentially
  declining SFR; e.g.][]{pforr12,dacunha15}, though they are still
analytic. Whatever the form of the continuous underlying SFH, it will
remain desirable to have a more physical (e.g. Gaussian or log-normal
distributed) description of the bursts of star formation, rather than
the top-hat models currently implemented.

One promising approach is that adopted by \citet{Pacifici12}, who
built half a million non-parametric SFHs by performing a semi-analytic
post-treatment of the Millenium simulation \citep{springel05}. The
downside of the increased complexity that we advocate is the increased
computation necessary for what is already a relatively load-intensive
task.\footnote{The latest version of the energy balance SED-fitting
  code {\sc Cigale} \citep{noll09} is not only parallelized, but
  also includes the particularly appealing capabilities of specifying
  arbitrary user-defined star formation histories; we intend to study
  its performance in a future investigation.} The desire for a more
varied set of SFHs can surely only increase as we embark upon the
survey era heralded by first light of e.g. the Large Synoptic Survey
Telescope \citep[][]{ivezic08} and the Square Kilometre
Array\footnote{www.skatelescope.org}, although the consistent
modelling and interpretation of these disparate data sets requires
considerable further investigation if we are to truly exploit their
immense potential \citep[e.g.][]{smith14}.

\subsection{The utility of the mock SED-based validation}

Having presented the results of our validation based on fitting the
SEDs of simulated galaxies, it is worthwhile to consider what this
controlled experiment tested. Because the simulations represent each
galaxy's stellar population as the sum of $\sim 10^5$ discrete
particles, the resulting SEDs reflect a diversity of ages and
metallicities, similar to real galaxies. Effects that make recovering
SFHs from SED modelling challenging include the fact that young stars
tend to dominate the luminosity at UV--optical wavelengths, thereby
obscuring older stellar populations \citep[see e.g.][for a recent
  discussion]{sorba15}; stellar isochrones change little at late
times, which makes it difficult to infer the shape of the early SFH
\citep[e.g.][]{bc03}; the true SFH can differ significantly from the
assumed parametric form (as discussed above); dust reddening is
degenerate with stellar age \citep[e.g.][]{gordon97}; and differential
obscuration can cause stars of different ages to be attenuated by
different amounts \citep[e.g.][]{charlot00}. All of these potential
barriers to SFH recovery are included in the simulations. It is thus
very encouraging that our method was able to recover the SFHs of the
simulated isolated disk galaxies relatively well, at least using the
median-likelihood SFHs.

\section{Conclusions} \label{S:conclusions}

We have presented modifications to the public version of the
\magphys\ SED-fitting code \citep[][]{dacunha08} which enable
statistical estimates of the star formation histories of individual
galaxies using photometric information alone (assuming that the
redshift is precisely known). Though \magphys\ is not intended for this purpose, our approach -- which uses the standard
\magphys\ stellar and dust SED libraries -- has been validated both
internally (by ``feeding" the code synthetic photometry corresponding
to an arbitrarily chosen SFH in the \magphys\ library) and
externally. Our external validation made extensive use of
state-of-the-art simulated ultraviolet to millimetre wavelength
photometry derived by performing three-dimensional dust radiative
transfer on SPH simulations from \citet[][]{Lanz14} using the
\sunrise\ code \citep[][]{Jonsson06,Jonsson10}. This approach to
validating SED fitting codes, which is discussed in detail in
\citet[][in which we highlighted how well \magphys\ can recover various properties of simulated galaxies, including the SFR, stellar mass and dust mass]{HS15}, gives us several advantages over real observations, and
enables us to test the efficacy of \magphys\ for recovering SFHs under
idealised conditions. Our main findings can be summarised as follows:

\begin{itemize}
	\item Using our modified version of \magphys, we are able to
          reliably recover the SFHs of isolated disk galaxies,
          provided that we marginalise over the library of
          SFHs. Marginalising over the libraries enables us to
          calculate median-likelihood SFHs in a manner analogous to
          how \magphys\ calculates galaxy parameters (e.g. stellar
          mass, dust luminosity) and naturally yields SFH
          uncertainties by estimating the percentiles of the SFH
          probability distribution functions as a function of
          look-back time.
	\item We find that SFHs corresponding to the best-fit of
          the \magphys\ model SEDs to the synthetic photometry are
          unreliable. This is manifest by large variations with
          viewing angle to the galaxy (the simulations include seven
          different viewing angles towards each model galaxy
          snapshot) and spurious bursty SFHs for galaxies which in
          truth have smoothly varying SFHs. The SFHs corresponding to
          the best photometric fit are a considerably worse estimate
          of the true SFH -- which is known for the simulations --
          than the median-likelihood SFH. We parametrise our SFH
          fidelity by introducing \chisqsfh\ (a goodness-of-fit
          comparing derived SFH estimates with the true values known
          from the simulation) and \dM\ (an estimate of the absolute mass
          differential between the true and modelled SFHs), which
          consistently favour the median-likelihood SFHs. We emphasise
          that neither of these parameters is used in the SED fitting
          itself, which is based purely on fitting model libraries to
          the synthetic photometry.
	\item We are unable to recover more complex SFHs, for example
          in the aftermath of a major merger-induced starburst,
          despite deriving a statistically acceptable fit to the
          photometric data (i.e. a reasonable $\chi^2$ goodness-of-fit
          parameter). This is particularly noteworthy given that in
          \citet{HS15}, we were able to reliably determine properties
          (such as stellar mass, star formation rate, specific star
          formation rate and dust luminosity) of a post-merger
          galaxy in this regime. It may be possible to better recover
          such complex SFHs by using SED templates based on SFHs
          extracted from semi-analytical models or cosmological simulations.
          \item The best-fit SFHs often contain spurious bursts, and even
          when there are bursts in the true SFHs, their properties (e.g.
          time of occurrence and duration) are not well recovered.
          Thus, one should interpret the relevant outputs, such as the stellar
          mass formed in bursts, with extreme caution.
\end{itemize}

To summarize, we recommend that the utmost care be exercised in the
interpretation of SFHs estimated from photometric data, and suggest that it is
essential to marginalise over a range of different possible SFHs if
any scientific value is required from their analysis (either studying
individual galaxies, or for example studying the contribution of
different galaxy samples to the evolving cosmic star formation rate
density). This caution should be further heightened if there is reason
to suspect a complex SFH (e.g. morphological tidal features, large far-infrared
luminosity, etc) unless an appropriate wide range of possibilities is
explicitly taken into account. 

\acknowledgments
The authors would like to thank the reviewer, Elisabete da Cunha, for an insightful report that improved the quality of this paper. 
We also wish to thank Charlie Conroy, Phil Hopkins, and Ben Johnson for useful discussions.
DJBS and CCH would like to acknowledge a financial award from the Santander Universities partnership scheme. 
CCH is grateful to the Gordon and Betty Moore Foundation for financial support and the University of Hertfordshire for hospitality.
DJBS would like the thank the Theoretical AstroPhysics Including Relativity and Cosmology (TAPIR) group at Caltech for their hospitality.
This research has made use of NASA's Astrophysics Data System Bibliographic Services.
\\

\footnotesize{

}

\begin{appendix}

\end{appendix}

\label{lastpage}

\end{document}